\documentclass{new_tlp}

\usepackage{amsmath}
\usepackage{amsfonts}
\usepackage{amssymb}
\usepackage{galois}
\usepackage{url}
\usepackage{tikz}
\usepackage{xspace}
\usepackage{multirow}
\usepackage{footnote}
\usepackage{stmaryrd}
\makesavenoteenv{tabular}
\makesavenoteenv{table}
\usepackage{amsmath}

\usepackage{listings}
\newcommand{\prettylstformat}[0]{
\lstset{language=Prolog,
        numbers=left,numberstyle=\tiny,stepnumber=1,numbersep=8pt,
        frameround=tttt,
        frame=ltrb,
        basicstyle=\scriptsize\ttfamily,
        commentstyle=\color{gray},
        breaklines=true,breakatwhitespace=true,
        showlines=true,
        showspaces=false,showtabs=false,
        keywords={pred,prop},
        escapeinside=~~,
      }}

\newcommand\costrelation{cost relation}
\newcommand\costrelations{cost relations}

\newcommand{\component}{cost center}

\newcommand{\nullenv}{\bot}
\newcommand{\env}{e}
\newcommand{\acc}{\predc} 
\newcommand{\changenv}{{\cal E}}
\newcommand{\environmentchange}{environment change}
\newcommand{\newapproach}{\textbf{New}}
\newcommand{\previousapproach}{\textbf{Prev}}
\newcommand{\extendedfeature}{$\times$}
\newcommand{\sameaccuracy}{$=$}

\newcommand{\benchmarks}{\textbf{Bench}} 
\newcommand{\accumcost}{\textbf{Acc. Cost}}
\newcommand{\compwprev}{\textbf{C}}
\newcommand{\statvsdyn}{\textbf{AvD}}
\newcommand{\timeunits}{s} 
\newcommand{\timeunitsfull}{seconds} 
\newcommand{\atime}{\textbf{Time (\timeunits)}} 
\newcommand{\stdcost}{\textbf{Std. Cost}} 
\newcommand{\numcalls}{\textbf{\#Calls}}
\newcommand{\bigo}{\textbf{Acc.BigO}}

\newcommand{\headcostbool}{B_\varphi} 
\newcommand{\headcost}{\varphi}

\newcommand{\litbool}{B} 
\newcommand\clausecost{$\headcost$ cost}

\newcommand\Prog{\mathcal P}

\newcommand\rs{{\cal R}_{\infty}}

\newcommand\cd{\Sigma}
\newcommand\ct{\vec \terme} 
\newcommand\cts{E}

\newcommand\szd{{\mathcal N}^{m}_{\top}}
\newcommand\sztypdom{\szd}

\newcommand\ruf[1]{{\mathcal C}_{#1}} 
\newcommand\eruf[1]{\hat{{\mathcal C}}_{#1}}  
  
\newcommand\ruccfsimp[2]{{\mathcal C}_{#1}^{#2}}

\newcommand\acalls{\leadsto_{\alpha}} 
\newcommand\artrcalls{\leadsto_{\alpha}^{\star}}
\newcommand\calls{\leadsto}

\newcommand\ccenters{\lozenge} 

\renewcommand\vec\bar

\newcommand\predp{\textup{\tt p}} 
\newcommand\predq{\textup{\tt q}} 
 
\newcommand\preds{\textup{\tt s}} 
\newcommand\predm{\textup{\tt m}} 
\newcommand\predw{\textup{\tt w}} 
\newcommand\predh{\textup{\tt h}} 
 
\newcommand\predc{\textup{\tt c}} 
 
\newcommand\predprime{\textup{\tt prime}}
\newcommand\predmultiple{\textup{\tt multiple}} 
\newcommand\predfact{\textup{\tt fact}} 
\newcommand\predprefix{\textup{\tt prefix}} 
\newcommand\predsuffix{\textup{\tt suffix}} 
\newcommand\predsublist{\textup{\tt sublist}}
\newcommand\predappend{\textup{\tt append}}

\newcommand\terme{\textup{\tt e}}

\newcommand\varx{\textup{\tt x}} 
\newcommand\vary{\textup{\tt y}} 
\newcommand\vecx{\vec \varx}
\newcommand\vecy{\vec \vary}

\newcommand{\ciao}{Ciao\xspace}
\newcommand{\ciaopp}{CiaoPP\xspace}

\def\imp{\hbox{${\tt \ :\!-\ }$}}             

\renewcommand\cd{{\Pi}} 

\newcommand\resr{r}

\newcommand\accube[4]{{\tt C}^{#2}_{#1,#3}(#4)}
\newcommand\accubena[3]{{\tt C}^{#2}_{#1,#3}}
\newcommand\acculbe[4]{{\tt C}^{#2}_{#1,#3}(#4)}
\newcommand\accub[3]{{\tt C}^{#2}_{#1}(#3)}
\newcommand\accubflops[3]{{\tt C}^{#1}_{#2}(#3)} 

\newcommand\stdub[2]{{\tt C}_{#1}(#2)}
\newcommand\stdubna[1]{{\tt C}_{#1}}
\newcommand\stdulb[2]{\stdub{#1}{#2}}

\newcounter{mnotei}
\newcommand{\mnote}[1]{%
  {\scriptsize\textsf{$^{\textcolor{red}{[n.\themnotei]}}$}}%
  \marginpar{\scriptsize\textsf{\textcolor{red}{[n.\themnotei]: #1}}}%
  \stepcounter{mnotei}
}
\renewcommand{\mnote}[1]{}

\newcommand{\optionaltext}[1]{#1}
\renewcommand{\optionaltext}[1]{ }

\newtheorem{example}{Example} 
\newtheorem{lemma}{Lemma} 

\title[Parametric Static Profiling]
      {A General Framework for Static Profiling of Parametric Resource Usage\ $^{\small 
\thanks{This research has received funding from 
EU FP7 agreement no 318337 \emph{ENTRA},
Spanish MINECO TIN2012-39391 \emph{StrongSoft} and 
TIN2015-67522-C3-1-R \emph{TRACES} projects, 
and the Madrid M141047003 \emph{N-GREENS} program. Special thanks are due
to John Gallagher for many fruitful and inspiring discussions and to
the anonymous reviewers for their detailed and useful comments.}}$}
       
\author[P.~L\'{o}pez-Garc\'{i}a et al.]{
  P. LOPEZ-GARCIA$^{1,2}$ ~~  M. KLEMEN$^{1}$ ~~ U. LIQAT$^{1}$ ~~ M.V. HERMENEGILDO$^{1,3}$ \\
\\ 
   $^1$IMDEA Software Institute \\
   \email{\{pedro.lopez,maximiliano.klemen,umer.liqat,manuel.hermenegildo\}@imdea.org} \\
   $^2$Spanish Council for Scientific Research  (CSIC) \\
   $^3$Technical University of Madrid (UPM) \\
   \vspace*{-1mm}
}

\jdate{October 2016}
\pubyear{2016}
\pagerange{\pageref{firstpage}--\pageref{lastpage}}

\submitted{April 30, 2016}
\revised{July 10, 2016}
\accepted{July 22, 2016}

\newcommand\beginequationspace{\vspace{-4mm}}
\newcommand\finishequationspace{\vspace{-3mm}}

\newcommand{\secbeg}{\vspace*{-3mm}}
\newcommand{\secend}{}

\newcommand{\betazero}{\beta}
\newcommand{\betaone}{\beta_1}
\newcommand{\betatwo}{\beta_2}

\renewcommand{\betazero}{n}
\renewcommand{\betaone}{n_1}
\renewcommand{\betatwo}{n_2}

\newcommand{\deltazero}{\delta}
\renewcommand{\deltazero}{n}

\newcommand{\variableV}{v}
\renewcommand{\variableV}{n}
\begin{document}

\label{firstpage}

\maketitle

\begin{abstract}

For some applications, 
standard 
resource analyses do not provide the information required.
Such analyses estimate the \emph{total} resource usage of a program (without
executing it) as functions on input data sizes. However, some
applications require knowing how such total
resource usage is \emph{distributed} over
selected parts of a program.
We propose a novel,
general, and flexible framework 
for setting up cost equations/relations
which can be instantiated for performing a wide range of resource
usage analyses, including both \emph{static profiling} and the
inference of the standard
notion of cost.
We extend and generalize 
standard
resource analysis techniques, 
so that the relations generated include additional Boolean control
variables
for switching on or off different terms in the
relations, as required by the desired resource usage profile. 
We also
instantiate our framework
to perform 
\emph{static profiling of accumulated cost}
(also parameterized by input data sizes).
Such information is much more useful to the software developer than
the 
standard
notion of cost: 
it identifies the parts of the program that have the greatest impact
on the total program cost, and which therefore should be optimized
first.
We also report on an implementation of our
framework within the \ciaopp system, and its instantiation for
accumulated cost, and provide some experimental results.
In addition to
generality,
our new method brings
important advantages over our previous
approach based on a program transformation, including support for
non-deterministic
programs, better and easier
integration in the compiler, and higher efficiency.

\end{abstract}

\begin{keywords}
 Static Profiling, Static Analysis, Resource Usage Analysis,
 Complexity Analysis 
\vspace*{-3mm}
\end{keywords}

\secbeg
\section{Introduction}
\label{sec:intro}
\secend

Resources are
numerical properties 
about the execution of a program, such as  
number of resolution steps, execution time, energy consumption, number
of calls to a particular
predicate,
number of network accesses, number of transactions in a database, etc.
The goal of automatic static cost analysis is estimating the resource usage 
of the execution of a program without running it, as a function of input
data sizes and possibly other (environmental) parameters.
The significant body of work on static analysis for logic
programs has actually also been applied to the analysis of other
programming paradigms, including imperative programs. 
This is achieved via a transformation of the
program into \emph{Horn
  Clauses}~\cite{decomp-oo-prolog-lopstr07-short}.
In this paper we concentrate on the analysis of Horn Clause programs,
independently of whether they are the result of a translation or the
actual program source.

Given a program $\Prog$ and a predicate $\predp \in \Prog$ of arity
$k$ and a set $\cd$ of $k$-tuples of
actual arguments
to $\predp$, we refer to the \emph{standard cost} of a call
$\predp(\ct)$ (i.e., a call to $\predp$ with actual data $\vec e \in
\cd$), as the resource usage (under a given cost metric) of the
complete execution of $\predp(\ct)$. Thus, the \emph{standard cost} is
a per-call cost formalized as a function $\ruf{\predp}: \cd
\rightarrow \rs$, where $\rs$ is the set of real numbers augmented
with the special symbol $\infty$ (which is used to represent
non-termination).
\emph{Standard cost}, and, in general, resource usage
information,
is very useful for a number of applications, such as automatic program
optimization, verification of resource-related specifications,
detection of 
performance bugs, or 
helping developers make resource-related design decisions. In the
latter case, the analysis has to show which parts of the program are
the most resource-consuming, i.e., which predicates 
would bring the highest overall improvement if they were optimized, so
that programming efforts can be focused more productively.  
The standard cost information only partially meets these objectives.
For example,
often predicates with the highest (standard) cost are not the ones
whose optimization is most profitable, 
since predicates which have lower costs but which are called more
often may be responsible
for a larger part of the overall resource usage. The input data sizes
to such calls are also relevant. Thus, rather than the global costs
provided by standard
cost analyses, 
what is really needed in many such applications 
is 
the results of a \emph{static profiling} of the program that 
helps identify 
the parts of a program responsible for 
highest 
fractions 
of the cost, or, more generally,
how the total resource usage of the execution of a program is
\emph{distributed} over
selected parts of it. 
By \emph{static profiling} we mean the static inference of 
the kinds of information that are
usually obtained at run-time by profilers.

For this reason, 
herein we are more interested in
what we refer to as \emph{accumulated cost}. To give an intuition of
this concept, we first explain our notion of \emph{cost centers},
which is similar to the one we use
in~\cite{staticprofiling-flops-short}, and was inspired
from~\cite{SansomJ95popl-short,morgan98}:
they are user-defined program points (predicates, in our case) to
which execution costs are assigned during the execution of a
program. Data about computational events is accumulated by the cost
center each time the corresponding program point is reached by the
program execution control flow.  Assume for example that
predicate $\predp$ calls another predicate $\predq$ (either directly
or indirectly), 
and that we declare that both predicates are cost centers.
In this case, the cost of a (single) call $\predp(\ct)$
\emph{accumulated in} cost center $\predq$, denoted
$\ruccfsimp{\predp}{\predq}(\ct)$, expresses how much of the standard
cost of $\predp(\ct)$ is attributed to $\predq$, and is the sum of the
costs of all the computations that are descendants (in the call stack)
of the call $\predp(\ct)$, and are performed ``under the scope'' of
any call to $\predq$.

We say that a computation is ``under the scope'' of a
call to cost center $\predq$, if the closest ancestor of
such computation in the call stack that is a cost center, is $\predq$. The
\emph{accumulated cost} is formalized as a function
$\ruccfsimp{\predp}{\predq}: \cd \rightarrow \rs$.
We refer the reader to~\cite{staticprofiling-flops-short} for a formal
definition of accumulated
cost.\footnote{In~\cite{staticprofiling-flops-short} we use the
  notation $\accubflops{\predp}{\predq}{\ct}$ instead of
  $\ruccfsimp{\predp}{\predq}(\ct)$.}

The goal of static analysis is to infer approximations (i.e.,
abstractions) of the concrete functions $\ruccfsimp{\predp}{\predq}$
and $\ruf{\predp}$ (or, more precisely, of the extensions of such functions to the
powerset of $\cd$) that represent the
\emph{accumulated} and \emph{standard} cost respectively.
In this paper we
propose a novel, general, and
flexible framework for setting up cost equations/relations which can
be instantiated for performing a wide range of static resource usage
analyses, including both 
\emph{accumulated cost} and
standard
cost.
Our starting point is the well-developed technique of setting up
recurrence relations representing resource usage functions
parameterized by input data
sizes~\cite{Wegbreit75-short-plus,Rosendahl89-short,granularity-short,caslog-short,low-bounds-ilps97-short,resource-iclp07-short,AlbertAGP11a-short,plai-resources-iclp14-short},
which are then solved to obtain (exact or safely approximated)
closed-forms of such functions (i.e., functions that provide upper or
lower bounds on resource usage in general).
\footnote{In addition, recently many other approaches have been proposed for
  resource
analysis~\cite{vh-03-short,DBLP:journals/toplas/0002AH12-short,grobauer01cost-short,igarashi02resource-short,nielson02automatic-short,DBLP:conf/ppdp/GieslSSEF12-short,conf-vmcai-AlbertGM11-short,DBLP:conf/popl/GulwaniMC09}.
  While based on different techniques, all these analyses are aimed at
  inferring the \emph{standard} notion of cost. Please
  see~\cite{staticprofiling-flops-short} for a further discussion of
  related work.}
Our proposal extends and generalizes these standard
resource analysis techniques 
by introducing into the derived relations extra Boolean control
variables whose value is 0 or 1.  A particular resource profile can be
analyzed by assigning values to the control variables, effectively
switching on or off different terms in the relations.  The 
standard
resource analysis is obtained by assigning 1 to all variables.
We also define a concrete Boolean variable assignment that
instantiates our framework so that it performs 
\emph{static profiling of accumulated cost}, similarly 
to~\cite{staticprofiling-flops-short}, where the results are also
parameterized by input data sizes.
However, the approach we present in this paper is 
quite different from our previous
approach~\cite{staticprofiling-flops-short}, 
which was based on a program transformation. The main
contributions of this paper and the differences and advantages
over that work
can be summarized as follows:

\vspace*{-2mm}
\begin{itemize}

\item We propose a novel, general, and flexible framework for setting
  up cost relations which can be instantiated for performing a wide
  range of resource usage analyses. 
  Is more general than~\cite{staticprofiling-flops-short},
  which is
  limited to accumulated cost analysis.

\item Our new approach can deal with
  non-deterministic/multiple-solution predicates, 
  unlike~\cite{staticprofiling-flops-short}. This is obviously a
  requirement for analyzing logic programs and is also useful for
  dealing with certain aspects of imperative programs, such as
  multiple dispatch;
  see~\cite{decomp-oo-prolog-lopstr07-short}. 
  While
  our previous approach could conceivably be extended to deal with
  such programs, it would certainly result in a more complicated and
  indirect solution.

\item Our new approach and its implementation are based on a direct
  application of abstract interpretation and integration into the
  \ciao\ preprocessor, \ciaopp~\cite{ciaopp-sas03-journal-scp-short},
  rather than on a program transformation.
  As a result, many useful \ciaopp{} features are
  inherited for free, such as \emph{multivariance} (being able to
  infer separate cost functions for different abstract call patterns
  for the same predicate), communication with the other required
  analyses, integrated treatment of special control features (such as,
  e.g., the cut), assertion-based verification and user interaction,
  efficient fixpoint,
  etc.
  Also, for this integration we define a novel abstract domain for
  resource analysis that keeps track of the \emph{environment}.

\item Furthermore, this direct implementation
  avoids
  the disadvantages of the transformation-based approach, such
  as making it more difficult to relate the results (and
  warnings/errors) to the original program, and complicating 
  the
  task of the auxiliary analyses needed for cost analysis (types,
  modes, determinism, non-failure, etc.). 
  This is because
  if the analyses
  are performed on the original program, then the results need to be
  transferred to the transformed program; and if the analyses are
  performed on the transformed program, then there is always the risk
  of loss of precision.
  Also, the transformation
  required by
  our previous approach is global, which is
  problematic for
  modular compilation.
  In general, this new approach allows much better and easier
  integration in a real-world compilation infrastructure.
  
\item The integration also inherits the capability of \ciaopp's
  analyzers
  of
  \emph{analyzing for several resources} at the same time.
  While it might be possible to define a new transformation
  capable of keeping track of several resources, this would further
  complicate the transformed program, and in any case requires
  additional work.

\item Finally, as our experimental results show,
  our new approach is more efficient than
  the transformation-based approach.  This is not only due to 
  its
  implementation as a direct abstract interpretation, but also to the
  inclusion and use of reachability information, performed
  automatically by the abstract interpretation framework.

\end{itemize}
\vspace*{-2mm}

\secbeg
\section{The 
Standard Parametric Cost Relations Framework}
\label{sec:starting-cost-re-framework}
\secend

We start by describing the kind of functions inferred by the standard
cost analysis that we generalize for static profiling.  Consider 
the function $\ruf{\predp}: \cd \rightarrow \rs$ introduced in the
previous section.
We extend
it to the powerset of $\cd$, i.e.,
$\eruf{\predp}: 2^{\cd} \rightarrow 2^{\rs}$, where
$\eruf{\predp}(\cts) = \{\ruf{\predp}(\ct) \mid \ct \in \cts\}$.  Our
goal is to abstract (safely approximate, as accurately as possible)
$\eruf{\predp}$
(note that $\ruf{\predp}(\ct) = \eruf{\predp}(\{\ct\})$). Intuitively,
this abstraction is the composition of two abstractions: a size
abstraction and a cost abstraction. The goal of the analysis is to
infer two functions $\eruf{\predp}^{\downarrow}$ and
$\eruf{\predp}^{\uparrow}: \sztypdom \rightarrow \rs$ that give lower
and upper bounds respectively on the cost function $\eruf{\predp}$,
where $\szd$ is the set of $m$-tuples whose elements are natural
numbers or the special symbol $\top$, meaning that the size of a given
term under a given size metric is \emph{undefined}.
Such bounds are given as a function of tuples of data sizes
(representing the concrete tuples of data of the concrete function
$\eruf{\predp}$). Typical size metrics are the actual value of a
number, the length of a list, the size (number of constant and
function symbols) of a term,
etc.~\cite{resource-iclp07-short,plai-resources-iclp14-short}.

Our starting point for static analysis is the standard
general framework described
in~\cite{resource-iclp07-short} for setting up parametric relations
representing the resource usage (and size relations) of programs and predicates.%
\footnote{
We give equivalent but simpler descriptions
  than in~\cite{resource-iclp07-short},
which are allowed by assuming that programs are the result of a
normalization process that makes all unifications explicit in the
clause body, so that the arguments of the clause head and the body
literals are all unique variables. We also omit the resource and
approximation identifiers, $\resr$ and $ap$ respectively, since they
are assumed to be arguments of all expressions that yield a resource
usage.}
The analysis infers size relations for each predicate in a program: 
arithmetic expressions that provide the size of output arguments of
the predicate as a function of its input data sizes. It also infers
size relations for each clause, which give the input data sizes of the
body literals as functions of the input data sizes to the clause
head.  Such size relations are instrumental for setting up 
\costrelations.
This work generalizes the approach
of~\cite{granularity-short,caslog-short,low-bounds-ilps97-short} to
infer \emph{user-defined resources} (by using an extension of the
\ciao assertion
language~\cite{hermenegildo11:ciao-design-tplp-short}).
The framework is doubly parametric: first, the costs inferred are
parametric (they are functions of input data sizes), and second, the
framework itself is parametric with respect to the 
resources being tracked
and 
the type of approximation made (upper or lower bounds).  Each
concrete resource $\resr$ to be tracked is defined by
two sets of (user-provided) functions, some of which can be constant
functions:

\vspace{-2mm}
\begin{enumerate}
\itemsep=0pt
\item \emph{Head cost 
$\headcost(P)$:} 
a function that returns the
  amount of resource $\resr$ used by the unification of the calling
  literal (subgoal) $P$ and the head of a clause matching $P$, plus
  any preparation for entering a clause (i.e., call and parameter
  passing cost).

\item \emph{Predicate cost} 
$\Psi(\predp)$: 
it is also possible
  to define the \emph{full cost} for a particular predicate $\predp$
  for resource $r$, i.e., the function 
$\Psi(\predp):\sztypdom \rightarrow \rs$ 
(with the sizes of $\predp$'s input data as
  parameters, $\vecx$) that returns the usage of resource $\resr$ made
  by a call to this predicate. This is specially useful for
  built-in
  or external predicates, i.e., predicates for
  which the source code is not available and thus cannot be analyzed,
  or for providing a more accurate function than analysis can
  infer.
  \footnote{Note that
    sometimes approximations have to be used when solving recurrence
    relations, and there are other potential sources of loss of
    precision
    in the
    intervening analyses,
    which can accumulate in larger programs. In these cases trust
    assertions can be used in key places to recover precision. While
    this implies a burden, it is certainly always better than having
    to do all the cost analysis of the program by hand.  }
$\Psi(\predp)$ is expressed using the \ciao assertion
  language ``trust'' assertions~\cite{hermenegildo11:ciao-design-tplp-short}. 
                                
\end{enumerate}

Thus, for a clause $C \equiv \predp(\bar{x}) \imp \predq_1(\bar{x}_1), \dots,
\predq_n(\bar{x}_n)$,
defining predicate $\predp$, the \costrelation{}
expressing the cost (for resource $r$) of the complete execution of a
single call to $\predp$ for 
input data sizes
$\vecx$ (obtaining all solutions),
represented as 
$\stdulb{\predp}{\vecx}$ 
is:

\beginequationspace
\begin{equation}\label{classrecurreq}
\stdulb{\predp}{\vecx} = \headcost(\predp(\vecx)) 
+ 
\sum_{i =1}^{lim(C,\vecx)} 
sols_i \times \stdulb{\predq_i}{\vecx_i} 
\end{equation}
\finishequationspace

\noindent
where 
$sols_i$ represents the product of the number of solutions produced by
the ancestor literals of $\predq_i(\bar{x}_i)$ in the clause body: 

\vspace*{-3mm}
\beginequationspace
\begin{equation}
sols_i = \prod_{j = 1}^{i-1}s_{pred}(\predq_j(\bar{x}_j))
\end{equation}
\finishequationspace

\noindent
where $s_{pred}(\predq_j(\bar{x}_j))$ gives the number of solutions
produced by $\predq_j(\bar{x}_j)$, and $lim(C,\vecx)$ gives the index of
the last body literal that is called in the execution of clause
$C$.

The (\emph{standard}) cost of a body literal $\predq_i(\bar{x}_i)$, i.e., 
$\stdulb{\predq_i}{\vecx_i}$, is obtained from the costs of all clauses
applicable to it that are executed, by using an aggregation operator
$\bigodot$.
The resulting set of \costrelations{} can be considered a definition
of the resource usage semantics of a program.  Ideally, we would like
to find solutions to such relations, i.e., closed-form functions that
give the resource usage of the programs and all of its
predicates. However, this is impossible to do statically for all 
cases, and we then seek approximations, both upper and lower bounds.
For this reason, we use a parametric operator $\bigodot(ap)$ that
depends on the approximation $ap$ being performed.
For example, if $ap$ is the identifier for lower bounds approximation
($lb$), then $\bigodot(ap)$ is the $min$ function.
If $ap$ is the identifier for upper bound approximation
($ub$), then a possible conservative definition for 
$\bigodot(ap)$ is the $\sum$ function. In this case, and since the
number of solutions generated by a predicate that will be demanded is
generally not known in advance, a conservative upper bound on the
computational cost of a predicate can be obtained by assuming that all
solutions are needed, and that all clauses are executed. Then, 
the cost of the predicate is assumed to be the sum of the costs of all of its
clauses. However, it is straightforward to take mutual exclusion into
account to obtain a more precise estimate of the cost of a predicate,
using the maximum of the costs of mutually exclusive groups of
clauses, as 
done in~\cite{plai-resources-iclp14-short}. 
Similarly, we use safe approximations of the function $lim(C, \vecx)$
in Expression~\ref{classrecurreq} by introducing the function
$lim(C,\vecx, ap)$ that returns 
the index of a literal in the clause body depending on the
approximation identifier $ap$.  For example, $lim(C,\vecx,ub) = n$
(the index of the last body literal) and $lim(C,\vecx,lb)$ 
is the index of the leftmost body literal that could
possibly fail.
\footnote{
  \ciaopp 
  implements analyses like coverage, non-failure, cardinality,
  reachability, modes, shapes, treatment of cut, etc.\ that are
  instrumental in this context;
  see~\cite{ciaopp-sas03-journal-scp-short} and its references.}
If the cost of a $\predq_i$ is given by a \emph{trust assertion} as a
function
$\Psi(\predq_i)(\vecy)$
then
the
closed-form 
$\Psi(\predq_i)(\vecx)$
is used directly instead of the symbolic
$\stdulb{\predq_i}{\vecx}$
appearing in the set of \costrelations{} to be solved.

\begin{example}
\label{examp:prime-numbers}

Consider the following program that checks whether a number $n$ is
prime based on Wilson's theorem: \emph{
  any integer $n > 1$ 
  is
  prime iff $(n-1)!\ \equiv\ -1 \pmod n$}. Equivalently, $n$ is prime
iff $(n-1)!\ + 1$ is a multiple of $n$.
\vspace*{-1mm}
\prettylstformat
\begin{lstlisting}[escapechar=\#]
prime(X):- X > 1, X1 is X - 1, #$\predfact$#(X1,F1), F is F1 + 1, multiple(F,X).

#$\predfact$#(X,1):- X = 1.
#$\predfact$#(X,Y):- X > 1, X1 is X - 1, #$\predfact$#(X1,Y1), Y is Y1*X.
\end{lstlisting}
\vspace*{-1mm}
\noindent
Assume that \predmultiple{} is a naively implemented library
predicate, so that its resource usage, in 
number of
resolution steps, is linear on the size of the input:
$\stdub{\predmultiple}{n,m} = n + 1$ if $n > 1$ (given by using a
trust assertion).
Assume that we want to infer the standard cost of this predicate in
resolution steps, i.e., we define $\headcost(\predp(\vecx))=1
\text{ for all predicates } \predp \in \Prog$. For brevity, we also
assume that we are only interested in inferring upper bounds on
resource usages, so that the expression $\stdub{\predp}{\vecx}$
appearing in Equation~\ref{classrecurreq} is understood to represent
an upper bound, and, assuming
no definite failure information, then 
$lim(C,\vecx)$ is the
index of the last body literal of clause $C$. Finally, we also assume that size
relations have been inferred for the different arguments in a clause,
and that the size metric used is the actual value of an argument,
since all arguments are numeric. Such relations are obvious in this
example, so that we focus only on \costrelations. However, as already
stated, 
\ciaopp{}
is able to infer and deal with a rich set of size metrics, and also
infer such size relations.  The size of the $kth$ output argument of
predicate $\predp$, given as a function of the input data sizes $\bar n$
to that predicate is represented as $Sz_{\predp}^{k}(\bar n)$.
It is important also to mention the modes of these predicates (again,
inferred automatically by \ciaopp): $\predprime$ has one input
argument and no output; $\predmultiple$ has two input arguments and no
output; and $\predfact$ has one input and one output, whose size we
have assumed is already inferred in terms of the size of the input by
the size analysis. This size is represented by
$Sz_{\predfact}^{2}(n)$, and is obtained from the setting up of the
following size relation:
\vspace{-2.5mm}
\begin{equation*}
    \begin{array}{rllrll}
      Sz_{\predfact}^{2}(n) = & 1 &  \text{if } n = 1 & \text{,}\quad
      Sz_{\predfact}^{2}(n) = & n \times Sz_{\predfact}^{2}(n-1) & \text{if } n > 1 \\ 
    \end{array}
\end{equation*}
\finishequationspace
\noindent 

\noindent
By solving this recurrence, the size analysis obtains the closed-form
$Sz_{\predfact}^{2}(n) = n!$.
Regarding the number of solutions, in this example all the predicates
generate at most one solution, thus $\forall i: sols_i=1$ in
Equation~\ref{classrecurreq}.
Now we have all the necessary elements to set up the \costrelations{} for
$\predprime$, $\predfact$, and $\predmultiple$:

\vspace*{-1mm}
\beginequationspace
\begin{equation*}
    \begin{array}{rll}
      \stdub{\predfact}{n} = & 1 &  \text{ if } n = 1 \\ [-1mm]
      \stdub{\predfact}{n} = & 1 + \stdub{\predfact}{n-1} & \text{ if
                                                            } n > 1 \\
      \\ [-4mm]
      \stdub{\predmultiple}{n,m} = & n + 1 & \text{ if } n > 1 
    \end{array}
\end{equation*}

\beginequationspace
\begin{equation*}
    \begin{array}{rll}
      \stdub{\predprime}{n} & = & 1 + \stdub{\predfact}{n-1}
                                  + \stdub{\predmultiple}{Sz_{\predfact}^{2}(n-1)
                                  + 1,n} \text{ if } n > 1 
    \end{array}
\end{equation*}
\finishequationspace

\vspace*{-2mm}
\noindent 
Note that in this program, the size of the input of the call to
$\predmultiple$ is given by the size of the output of $\predfact$,
represented by $Sz_{\predfact}^{2}(n)$. 
After solving these equations and composing the closed forms, we
obtain the following closed form functions:

\vspace*{-1mm}
\beginequationspace
\begin{equation*}
    \begin{array}{rll}
      \stdub{\predfact}{n} = & n &  \text{ if } n > 1 \\ [-1mm]
      \stdub{\predmultiple}{n,m} = & n + 1 & \text{ if } n > 1 \\ \\ [-3mm]
      \stdub{\predprime}{n} = & (n - 1)!+n+3 & \text{ if } n > 1   \\
    \end{array}
\end{equation*}
\finishequationspace
\end{example}

\vspace*{-5mm}
\begin{example}
\label{examp:standard-cost}
Consider the following program $\Prog$:

\begin{minipage}{0.4\textwidth}
\prettylstformat
\begin{lstlisting}
p(X,Y):- h(X), q(X,Y), w(Y), s(X).

q(0,_).
q(X,Y):- X > 0, X1 is X - 1, 
         m(Y), q(X1,Y), s(X).
\end{lstlisting}
\end{minipage}
\hspace*{6mm}
\begin{minipage}{0.47\textwidth}
\vspace*{-7mm}
\prettylstformat
\begin{lstlisting}[firstnumber=8]
m(0).
m(X):- X > 0, w(X), X1 is X - 1, m(X1).

s(0).
s(X):- X > 0, X1 is X - 1, w(X), s(X1).

h(2).
h(3).
\end{lstlisting}
\end{minipage}

\noindent
Assume as in the previous example that we want to infer upper bounds
of the standard costs of all the predicates in resolution steps, i.e.,
$\headcost(\predp(\vecx))=1 \text{ for all predicates } \predp \in
\Prog$.  
Assume also that $\predw$ is a library predicate and that its (standard)
cost is given as a predicate cost function (by using a trust
assertion): 

\vspace*{-2mm}
\beginequationspace
\begin{equation}
\label{trust-for-w}
\Psi(\predw)(n) = 2 n + 1
\end{equation}
\finishequationspace

\vspace*{-2mm}
\noindent
We assume again that the size metric used is the actual value of the
arguments, since
they are all numeric, and that size relations, again obvious, have
been inferred for
all clause arguments, 
which are all inputs, 
and we focus only on \costrelations. 
The \costrelation{} for the recursive
clause of predicate $\preds$, according to
Expression~\ref{classrecurreq} is 
(for simplicity, $sols_i =1$ for all predicates in this example):

\vspace*{-2mm}
\beginequationspace
\begin{equation*}
    \begin{array}{rll}
      \stdub{\preds}{n} = & 1 + \stdub{\predw}{n} + \stdub{\preds}{n-1} & \text{if } n > 0
    \end{array}
\end{equation*}
\finishequationspace

\vspace*{-2mm}
\noindent
Since $\stdub{\predw}{n}$ is given by a trust assertion as
$\Psi(\predw)(n) = 2 \ n + 1$ (Expression~\ref{trust-for-w}), this
\costrelation, together with the one for the
non-recursive clause, form the system:

\vspace*{-1mm}
\beginequationspace
\begin{equation*}
    \begin{array}{rll}
      \stdub{\preds}{n} = & 1 &  \text{if } n = 0 \\
      \stdub{\preds}{n} = & 1 + 2 n + 1 + \stdub{\preds}{n-1} & \text{if } n > 0
    \end{array}
\end{equation*}
\finishequationspace

\noindent 
and its closed-form solution is 
$\stdub{\preds}{n} = n^2+3n + 1 \text{ for } n \geq 0$. 
The same \costrelations{} correspond to predicate $\predm$, therefore
its closed form is $\stdub{\predm}{n} = n^2+3n+1 \text{ for } n \geq
0$.  For predicate $\predh$, the following 
non-recursive system of \costrelations{} is set up:

\vspace*{-2mm}
\beginequationspace
\begin{equation*}
    \begin{array}{cccccccc}
      \stdub{\predh}{n} = 1, \text{ if } n = 2 & 
      & & & \text{ and } & & & 
      \stdub{\predh}{n} = 1, \text{ if } n = 3
    \end{array}
\end{equation*}
\finishequationspace
\vspace*{-2mm}

\noindent
obtaining $\stdub{\predh}{n} = 1$, since the clauses of $\predh$ are
mutually exclusive.
Now, the \costrelations{} for $\predq$ are:

\vspace*{-6mm}
\beginequationspace
\begin{equation*}
    \begin{array}{rll}
      \stdub{\predq}{m,n} = & 1 & \text{if } m = 0
      \\ \stdub{\predq}{m,n} = & 1 + \stdub{\predm}{n} +
      \stdub{\predq}{m -1,n} + \stdub{\preds}{m} & \text{if }
      m > 0
    \end{array}
\end{equation*}
\finishequationspace

\vspace*{-1mm}
\noindent
Replacing $\stdub{\predm}{n}$ and $\stdub{\preds}{n}$ with their
corresponding closed-form functions obtained before, and solving the
recurrence, we obtain
$\stdub{\predq}{m,n} = \frac{1}{3}m^3+mn^2+2m^2+3mn+\frac{14}{3}m+1$.
Finally, the \costrelations{} for the main predicate $\predp$ result
in:

\vspace*{-2mm}
\beginequationspace
\begin{equation*}
    \begin{array}{rll}
      \stdub{\predp}{m,n} = & 1 + \stdub{\predh}{m} + \stdub{\predq}{m,n} + \stdub{\predw}{n} + \stdub{\preds}{m} &  \\
    \end{array}
\end{equation*}
\finishequationspace

\vspace*{-2mm}
\noindent 
and its closed form is: 
$\stdub{\predp}{m,n} =
\frac{1}{3}m^3+mn^2+3m^2+3mn+\frac{23}{3}m+2n+4$.
\end{example}
\vspace{-0.3cm}

\secbeg
\section{Generalizing the 
Standard
Cost Relations Approach}
\label{sec:generalization}
\secend

Our proposal extends and generalizes the approach described in
Sect.~\ref{sec:starting-cost-re-framework}.
We introduce a new concept of cost, 
$\acculbe{\predp}{\acc}{\env}{\vecx}$, representing the (part of the)
cost of the complete execution of a single call 
$\predp(\vecx)$ (i.e.,
$\stdulb{\predp}{\vecx}$ in
Sect.~\ref{sec:starting-cost-re-framework}), performed in an 
\emph{environment} $\env$, that is attributed/assigned to
\emph{\component}
$\acc$ of the program.  
The parameter $\env$ is used to capture a broad notion of \emph{environment}.
For example,
it can be just the name of a
predicate that is an ancestor of $\predp$ in the call stack.
In a more complex setting,
for example when inferring hardware-dependent resources, such as energy
~\cite{NMHLFM08-tooshort,isa-energy-lopstr13-final-short,isa-vs-llvm-fopara-short},
$\env$ can also include information about the state of the hardware
(or the whole system, including the running software environment), 
e.g., the last instruction executed (
useful for modeling the
\emph{switching cost} of instructions), temperature, voltage,
cache state, and pipeline state.
There is of course a trade-off between the amount of information
in  $\env$ and analysis efficiency and accuracy.

As already said, and similarly to~\cite{staticprofiling-flops-short},
in this paper we assume that a \emph{cost center} is a predicate in the
program.  Conceptually, we can say that we extend the notion of
resource so that it is now
a pair $(\acc, \resr)$, where $\resr$
is a resource identifier as before (e.g., resolution steps, execution
time, energy, etc.), and $\acc$ is the \component{} (predicate) 
that the resource usage is attributed/assigned to.

We also introduce \emph{Boolean functions} $\headcostbool(\predp, \acc,
\env)$
and $\litbool(\predp, \acc, \env, \predq)$ to control which terms of the
\costrelation{} should be 
considered. To this end,
Expression~\ref{classrecurreq} is generalized as: 

\vspace*{-2mm}
\beginequationspace
\begin{equation}\label{genrecurreq}
\begin{array}{rl} 
\acculbe{\predp}{\acc}{\env}{\vecx} & = \headcostbool(\predp,\acc,\env)
\times \headcost(\predp(\vecx)) + \sum_{i =1}^{lim(C,\vecx)} sols_i \times
\litbool(\predp,\acc,\env, \predq_i) \times
\acculbe{\predq_i}{\acc}{\env'}{\vecx_i}
\end{array}
\end{equation}
\finishequationspace

\vspace*{-2mm}
\noindent
where $\env'= \changenv(\predp,\acc,\env, \predq_i(\vecx_i))$, and $\changenv$
is the \emph{\environmentchange} function, which obtains the new
environment for $\predq_i$.
If the cost of $\predp$ is given (by using a trust assertion) as a
function $\Psi(\predp)(\vecx)$,
then: 

\vspace*{-6mm}
\beginequationspace
\begin{equation}
\label{eq:trustcost}
\acculbe{\predp}{\acc}{\env}{\vecx} = \headcostbool(\predp,
\acc, \env) \times \Psi(\predp)(\vecx)
\end{equation}
\finishequationspace

Again, this equational framework can be instantiated to obtain the standard
cost by defining 
$\headcostbool(\predp, \acc, \env)
= \litbool(\predp, \acc, \env, \predq) \equiv 1$, and defining
$\changenv$ so that it does not change the environment and always
returns the input environment, i.e., $\changenv(\predp,\acc,\env,\predq_i(\vecx_i)) = \env$.
The \emph{standard} cost
$\stdulb{\predp}{\vecx}$ is then given by 
$\acculbe{\predp}{\predp}{\nullenv}{\vecx}$, where $\nullenv$ is the
\emph{null} environment, in which no information about the environment
is tracked, and the only \component{} that the cost of a single call
to $\predp$ is attributed to is the predicate $\predp$ itself.

\secbeg
\section{Instantiation for Parametric Accumulated-cost Static Profiling}
\label{sec:accumulated-cost-inst}
\secend

We now instantiate the general approach described in
Sect.~\ref{sec:generalization} for the static inference of accumulated
cost.
The advantages of this approach with respect to our previous approach
to accumulated cost inference~\cite{staticprofiling-flops-short} were
already discussed in Sect.~\ref{sec:intro}.

Assume we are given a set of (user-defined) cost centers $\ccenters$,
which, as mentioned before, in our approach are
program predicates.
Assuming that $\predp$ is a cost center, the standard cost of a single
call $\predp(\vecx)$ (as defined in Sect.~\ref{sec:intro}, and whose
inference was discussed in
Sect.~\ref{sec:starting-cost-re-framework}) is the 
sum of its accumulated costs in all the cost centers in the program, 
or, equivalently in all the cost centers that are descendants (in the call stack) of $\predp$.
This is formally expressed in~\cite{staticprofiling-flops-short}
Theorem 1, and, intuitively, the proof is based on the fact that,
according to the definition of accumulated cost, the cost of any
computation performed during the complete execution of $\predp(\vecx)$
is uniquely attributed to a cost center (predicate): the closest
ancestor of such computation in the call stack that is a cost center.

Given a
predicate $\predp$, we refer to the computations performed by
a call $\predp(\vecx)$ that are not under the scope of any cost center
that is a descendant (in the call stack) of $\predp$, as the
\emph{residual computations} of $\predp$.  We refer to the cost of
such computations as the \emph{residual cost} of $\predp$. Note that
such computations include the computations performed by calls to 
non-cost-center predicates that are descendants of $\predp$ and that
are not under the scope of any cost center that is a descendant of $\predp$.
Assume that the analysis is inferring accumulated costs on a given
cost center $\predc$.
When analyzing a call to a non-cost-center predicate $\predp$, 
its residual cost must be attributed to $\predc$ only if the
call $\predp(\vecx)$ is under the scope of $\predc$ (i.e., is a
descendant of $\predc$).
When analyzing a call to a cost-center predicate $\predp$, 
its residual cost must be attributed to $\predc$ only if $\predp = \predc$.
Thus, in the expression
$\acculbe{\predp}{\predc}{\env}{\vecx}$ (where necessarily $\predc \in
\ccenters$) the environment $\env$
is just a Boolean value representing whether the (single) call to
$\predp$ is in the scope of cost center $\predc$ ($\env = 1$) or not
($\env = 0$).

To this end, we define the \emph{\environmentchange} function as follows:
$\changenv(\predp,\acc,\env, \_) \equiv (\predp = \acc \lor (\predp \not\in
\ccenters \land \env))$.

Knowing that a given predicate cannot be called by another 
during program execution allows the analysis to ignore some parts not
affecting the cost to be inferred. 
We define a simple \emph{calls} relation between predicates
as:
$\predp$ calls $\predq$, denoted $\predp \acalls \predq$, if
and only if a literal with predicate symbol $\predq$ appears in the
body of a clause defining $\predp$; 
$\artrcalls$ is the reflexive transitive closure of $\acalls$. 
This $\acalls$ relation is an
abstraction (over-approximation) of the concrete $\calls$ relation 
(a more precise abstraction is computed by \ciaopp). 

The Boolean assignment functions (appearing in
Expression~\ref{genrecurreq}) are defined as
follows:

\vspace*{-2mm}
\beginequationspace
\begin{equation}
\label{eq:headcostbool}
\headcostbool(\predp, \acc, \env) \equiv (\predp = \acc \lor (\predp \not\in \ccenters \land \env ) )
\end{equation}
\finishequationspace

\vspace*{-1mm}
\beginequationspace
\beginequationspace
\begin{equation}
\label{more-efficient-litbool}
\litbool(p, \acc, \env, q) \equiv \headcostbool(p, \acc, \env) \lor (q \artrcalls \acc)
\end{equation}
\finishequationspace
\vspace*{-2mm}

Note that the analysis of the accumulated cost of a given
non-cost-center predicate $\predp$ in a given cost center $\acc$ can
create at most two versions of $\acculbe{\predp}{\acc}{\env}{\vecx}$
for the same input (calling pattern) $\vecx$ 
(and hence, there
will be at most two versions of the \costrelations{} for $\predp$): the
version $\acculbe{\predp}{\acc}{1}{\vecx}$ created if there is a
(direct or indirect) call to $\predp$ in the scope of $\acc$, e.g., if
such call is in the body of a clause defining $\acc$ (in which case
the \emph{\clausecost} is added to the \costrelations{} for $\predp$),
and the variant $\acculbe{\predp}{\acc}{0}{\vecx}$ created if there
is a call to $\predp$ not in the scope of $\acc$
(in which case the \emph{\clausecost} is not added).

\vspace*{-2mm}
\begin{lemma}
\label{lem:noenvi-bool}
$\forall \predp, \predq \in \ccenters$, $\forall \env \in \{0,1\}$, it holds that
$\changenv(\predp,\predq,\env, \_) \equiv (\predp = \predq)$ and
$\headcostbool(\predp,\predq,\env) \equiv (\predp = \predq)$. 
\end{lemma}
\noindent
This implies that:

\vspace*{-2mm}
\begin{lemma}
\label{lem:noenvironment}
$\forall \predp, \predq \in \ccenters$ it holds that
$\accube{\predp}{\predq}{0}{\vecx} =
\accube{\predp}{\predq}{1}{\vecx}$.
\end{lemma}

\noindent
Thus,
if $\predp \in \ccenters$ we omit the environment $\env$ and
write $\accub{\predp}{\predq}{\vecx}$. Note that necessarily $\predq
\in \ccenters$.

\vspace*{-2mm}
\begin{lemma}
\label{lem:zerocost}
$\forall \predp, \predq \in \ccenters$,  if $\predp \not\artrcalls \predq$ then 
$\accub{\predp}{\predq}{\vecx} = 0$.
\end{lemma}

\vspace*{-2mm}
\begin{lemma}
\label{lem:zerocost-non-cost-center}
$\forall \predp \not\in \ccenters, \forall \predq \in \ccenters$,  if $\predp \not\artrcalls \predq$ then 
$\accube{\predp}{\predq}{0}{\vecx} = 0$.
\end{lemma}

\vspace*{-2mm}
Note also that in the standard \costrelation-based static analysis,
\costrelations{} are set up for each predicate in the program.  In the
approach we propose here for accumulated cost, \costrelations{} are
set up for each cost center and for each predicate in the program.

\begin{example}
\label{examp:prime-numbers2}
In Example \ref{examp:prime-numbers}, predicate
$\predprime$ was found too expensive in terms of resolution steps to be
practical, since $\stdub{\predprime}{n} \in \mathcal{O}(n!)$. However,
the standard cost inferred for all the predicates
called from
$\predprime$ is linear, and it is not easy to detect at first glance
where the resource is really consumed.
To locate the culprit, 
traditionally this would be attempted using a dynamic
profiling tool, executing the program with several test cases 
--commonly known as \emph{hot spot detection}. However, as
with the standard cost analysis, we want to detect such hot spots
\emph{statically}, in order to have sound information for \emph{any}
possible input. For this purpose,
we perform the accumulated cost analysis
declaring that all predicates are cost centers
(i.e,
$\ccenters = \{\predprime,\predfact,\predmultiple\}$). 
Based on 
the equational framework instantiation above
and Lemma~\ref{lem:noenvironment}, consider the
cost of a single call to $\predprime$ 
accumulated in $\predfact$,  $\accub{\predprime}{\predfact}{n}$, for an input size $n$. 
As already stated, the number of solutions of all these
predicates is 1, and the output sizes have already been inferred. For
the sake of conciseness, from now on we refer to $\predprime$,
$\predfact$ and $\predmultiple$ as $p$, $f$ and $m$ respectively. The \costrelations{} for 
the accumulated costs in cost center $\predfact$ are:  

\vspace*{-2mm}
\beginequationspace
\begin{equation*}
    \begin{array}{rll}
      \accub{p}{f}{n} = & \headcostbool(p,f,\_) \times \headcost(p(n)) +  
      \accub{f}{f}{n-1} + \accub{m}{f}{Sz_{f}^{2}(n-1) + 1} & 
    \end{array}
\end{equation*}
\vspace*{-3mm}
\begin{equation*}
    \begin{array}{rll}
      \accub{f}{f}{n} = & \headcostbool(f,f,\_) \times \headcost(f(n)) & \text{if } n = 1 \\  
      \accub{f}{f}{n} = & \headcostbool(f,f,\_) \times \headcost(f(n)) + \accub{f}{f}{n-1}  & \text{if } n > 1   
    \end{array}
\end{equation*}
\vspace*{-3mm}
\begin{equation*}
    \begin{array}{rll}
      \accub{m}{f}{n} = & \headcostbool(m,f,\_) \times \Psi(m)(n) = \headcostbool(m,f,\_) \times (n+1)  &  \text{if } n > 1   
    \end{array}
\end{equation*}
\finishequationspace

\vspace*{-2mm}
\noindent
We have that $\headcost(\_) = 1$, and, according to
Expression~\ref{eq:headcostbool}, $\headcostbool(p,f,\_) =
\headcostbool(m,f,\_) = 0$ and $\headcostbool(f,f,\_) = 1$.
Using these values, the \costrelations{} defining 
$\accub{p}{f}{n}$ are:

\vspace*{-1mm}
\beginequationspace
\begin{equation*}
    \begin{array}{rll}
      \accub{p}{f}{n} = & \accub{f}{f}{n-1} &  
    \end{array}
\end{equation*}
\vspace*{-4mm}
\begin{equation*}
    \begin{array}{rll}
      \accub{f}{f}{n} = & 1 & \text{if } n = 1 \\  
      \accub{f}{f}{n} = & 1 + \accub{f}{f}{n-1}  & \text{if } n > 1   
    \end{array}
\end{equation*}
Solving this system of \costrelations, we finally obtain: \ \ \  $\accub{p}{f}{n} = n$.
\noindent
Analogously, we obtain the closed-form functions for $\accub{p}{m}{n}$ and $\accub{p}{p}{n}$:

\beginequationspace
\begin{equation*}
    \begin{array}{rll}
      \accub{p}{p}{n} = & 1 & \text{if } n > 1 \\ 
      \accub{p}{m}{n} = & (n-1)! + 2 & \text{if } n > 1 \\
    \end{array}
\end{equation*}
\finishequationspace

\noindent
Now, it is clear that the most expensive part of this program is the
call to \predmultiple.
Even though the standard cost
of 
\predmultiple{}
is 
linear, its input size is the
output size of the call to $\predfact$ (plus 1), 
which is 
the factorial of the input to \predprime{} minus 1.
In this case the problem can really only be fixed by using a better
implementation of \predmultiple{} ($\mathcal{O}(1)$) or of 
\predprime,
to achieve the expected polynomial resource usage.

This example illustrates how the accumulated cost 
is more useful than 
the standard cost. 
Neither the standard cost of $\predmultiple$ ($n+1$) nor the number of
calls to this predicate from $\predprime$ (since it is called just
once) gives a direct hint that this predicate is responsible for most of
the resource consumption of $\predprime$.
\end{example}

\vspace*{-3mm}
\begin{example}
\label{examp:accum-cost}
Consider again the program in Example~\ref{examp:standard-cost}.  Assume
that we declare that predicates 
$\predp$, $\predq$, $\predm$ and $\predh$ 
are cost centers, i.e., $\ccenters =
\{\predp, \predq,\predm, \predh \}$, 
and 
$\preds$ and $\predw$ are not. 
For space reasons, we will only illustrate the inference of upper
bounds on accumulated costs in all cost centers. 

The accumulated costs in cost center $\predq$ are inferred as follows.
Consider the clause defining predicate $\predp$.
Since $\predp \in \ccenters$, by Lemma~\ref{lem:noenvi-bool} 
the current environment $\env$ is irrelevant for the computation of
the new environment $\env'$ (i.e., $\env' =
\changenv(\predp,\predq,0,\_) = \changenv(\predp,\predq,1,\_) \equiv
(\predp = \predq) \equiv 0$),
and for the computation of the head cost, i.e.,
$\headcostbool(\predp,\predq,0) = \headcostbool(\predp,\predq,1)
\equiv (\predp = \predq) \equiv 0$.  
Thus, the \costrelation{} for $\predp$ according to
Equation~\ref{genrecurreq} is
$\accub{\predp}{\predq}{\varx,\vary} = \accub{\predh}{\predq}{\varx} +
\accub{\predq}{\predq}{\varx,\vary} +
\accube{\predw}{\predq}{0}{\vary} +
\accube{\preds}{\predq}{0}{\varx}$.
Consider predicate $\predq$ now. 
Since $\headcostbool(\predq,\predq, \env) \equiv (\predq = \predq)
\equiv 1 
\text{ and } \changenv(\predq,\predq,e,\_) \equiv (\predq = \predq) \equiv 1
\text{ for } \env \in \{0,1\}$, the \costrelations{} for the base case
and recursive clause of $\predq$ respectively are:

\beginequationspace
\begin{equation*}
\begin{array}{ll} 
\accub{\predq}{\predq}{\varx,\vary} = \headcostbool(\predq,\predq, \_) \times 1 = 1 \times 1 = 1 & \text{if } \varx = 0 \\
\accub{\predq}{\predq}{\varx,\vary} = 1 + \accub{\predm}{\predq}{\vary}
+ \accub{\predq}{\predq}{\varx-1,\vary} + \accube{\preds}{\predq}{1}{\varx} & \text{if } \varx > 0
\end{array}
\end{equation*}
\finishequationspace

\noindent
For expression $\accube{\preds}{\predq}{1}{\varx}$ appearing in
the recursive \costrelation{} for $\predq$ above (i.e., the version of the cost
of $\preds$ when
called in the scope of cost center $\predq$),
the \costrelations{} are:\footnote{Since $\preds \not\in \ccenters$,
  the environment is needed in this case.}

\vspace*{-1mm}
\beginequationspace
\begin{equation*}
\begin{array}{lll} 
\accube{\preds}{\predq}{1}{\varx} = & 1 & \text{if } x = 0 \\
\accube{\preds}{\predq}{1}{\varx} = & 1 + \accube{\predw}{\predq}{1}{\varx} + \accube{\preds}{\predq}{1}{\varx-1} & \text{if } \varx > 0 \\
\end{array}
\end{equation*}
\finishequationspace

\noindent
We now need to infer the function represented by expression
$\accube{\predw}{\predq}{1}{\varx}$ appearing in the recursive
\costrelation{} for $\preds$ above.  Since the cost function for
$\predw$ is given by a trust assertion (see
Expression~\ref{trust-for-w}) and $\headcostbool(\predw, \predq, 1) =
1$, we have that $\accube{\predw}{\predq}{1}{\varx} =
\headcostbool(\predw, \predq, 1) \times (2 \ \varx + 1) = 2 \ \varx +
1$.  Using this function, the closed-form solution for
$\accube{\preds}{\predq}{1}{\varx}$ is $\varx^2 + 3\varx + 1 \text{
  for } \varx \geq 0$.
For expression $\accube{\predw}{\predq}{0}{\vary}$ appearing in
the equation for $\predp$ above, we have that
$\accube{\predw}{\predq}{0}{\vary} = \headcostbool(\predw, \predq, 0)
\times (2 \vary + 1) = 0 \times (2 \vary + 1) = 0$.
Now, for expression $\accube{\preds}{\predq}{0}{\varx}$ appearing
in the \costrelation{} for $\predp$ above (i.e., the version of the
cost of $\preds$ when it is called not in the scope of cost center
$\predq$),
we have that $\accube{\preds}{\predq}{0}{\varx} = 0$
(Lemma~\ref{lem:zerocost-non-cost-center}).
For expression $\accub{\predm}{\predq}{\vary}$ appearing in the
second \costrelation{} for $\predq$ above, we have that
$\accub{\predm}{\predq}{\vary} = 0$ (Lemma~\ref{lem:zerocost}), 
and no \costrelation{} is set up for predicate $\predm$.
Now, the accumulated costs in cost center $\predh$ are inferred as follows.
The accumulated cost in $\predh$ for a call to $\predp$ is given by:

\vspace*{-1mm}
\beginequationspace
\begin{equation*}
\accub{\predp}{\predh}{\varx,\vary} = \accub{\predh}{\predh}{\varx} +
\accub{\predq}{\predh}{\varx,\vary} +
\accube{\predw}{\predh}{0}{\vary} + \accube{\preds}{\predh}{0}{\varx}
\end{equation*}
\finishequationspace
\vspace*{-1mm}

\noindent
We have that:
\vspace*{-2mm}
\beginequationspace
\begin{equation*}
\accub{\predq}{\predh}{\varx,\vary} = 0 \text{ and }
\accub{\predm}{\predh}{\vary} = 0 \text{ (by Lemma~\ref{lem:zerocost})}
\end{equation*}
\finishequationspace
\noindent
and: 
\beginequationspace
\begin{equation*}
\accube{\preds}{\predh}{0}{\varx} = 0  \text{ (by Lemma~\ref{lem:zerocost-non-cost-center})} 
\end{equation*}

\noindent
and $\accube{\predw}{\predh}{0}{\vary} = \headcostbool(\predw,\predh,
0) \times  \Psi(m)(\vary) = 0$.  Then, the \costrelations{} for the
accumulated cost in $\predh$ for a call to $\predh$ are:
\vspace*{-2mm}
\beginequationspace
\begin{equation*}
\begin{array}{ll}
\accub{\predh}{\predh}{\varx} = & \headcostbool(\predh,\predh, \_)
\times 1  
= 1
\\ \accub{\predh}{\predh}{\varx} = & \headcostbool(\predh,\predh, \_)
\times 1 = 1
\end{array}
\end{equation*}
\finishequationspace

\vspace*{-1mm}
\noindent
Therefore, $\accub{\predh}{\predh}{\varx} = 1$ and
$\accub{\predp}{\predh}{\varx,\vary} = 1$.
For cost center $\predm$ we have:

\beginequationspace
\begin{equation*}
\begin{array}{ll}
\accub{\predp}{\predm}{\varx, \vary} =  \varx \vary^2 + 3 \varx \vary + 2 \vary + \frac{1}{3} \varx^3 + \frac{5}{2} \varx^2 + \frac{25}{6} 
\varx + 1 & \\ [1mm]
\accub{\predq}{\predm}{\varx, \vary} =  \varx \vary^2 + 3 \varx \vary + \frac{1}{3} \varx^3 + \frac{3}{2} \varx^2 + \frac{13}{6} 
\varx & \\ [1mm]
\accub{\predm}{\predm}{\varx} =  \varx^2 + 3 \varx + 1 &
 \\ 
\end{array}
\end{equation*}
\finishequationspace

\noindent
Finally, for cost center $\predp$ we have:

\beginequationspace
\begin{equation*}
\begin{array}{cccc}
\accub{\predp}{\predp}{\varx, \vary} = 1 &
\accub{\predq}{\predp}{\varx, \vary} = 0 & 
\accub{\predm}{\predp}{\varx} = 0 & \accub{\predh}{\predp}{\varx} = 0 \\ [1mm]
\end{array}
\end{equation*}
\finishequationspace

\vspace*{-3mm}
\noindent 
Note that the large complexity of $\accub{\predp}{\predm}{\varx, \vary}$
makes us realize that if we move the call $\predm(\vary)$ from the
recursive clause of $\predq$ to the clause of $\predp$:
\prettylstformat
\begin{lstlisting}
p(X, Y):- h(X), m(Y), q(X, Y), w(Y), s(X).

q(0, _).
q(X, Y):- X > 0, X1 is X - 1, q(X1, Y), s(X).
\end{lstlisting}

\noindent
then, the standard cost of $\predp$ will be reduced. In particular, it
is reduced from $\stdub{\predp}{\varx,\vary} =
\frac{1}{3}\varx^3+\varx\vary^2+3\varx^2+3\varx\vary+\frac{23}{3}\varx+2\vary+4$
to $\stdub{\predp}{\varx,\vary} = \vary^2 + 5 \vary +
\frac{1}{3}\varx^3 + 3 \varx^2 + \frac{20}{3} \varx + 8$.

\vspace*{-2mm}
\end{example}

\secbeg
\section{Implementation and Experimental Results}
\label{sec:exp-results}
\secend

We have implemented the proposed approach within the \ciaopp system, by
extending the implementation of~\cite{plai-resources-iclp14-short}. The
latter improved on~\cite{resource-iclp07-short} by defining the resource
analysis itself as an \emph{abstract domain} that is integrated into
the PLAI abstract interpretation framework~\cite{ai-jlp-short,inc-fixp-sas-short}
of \ciaopp, inheriting features such as multivariance, efficient
fixpoints, and assertion-based verification and user interaction. A
significant additional improvement brought about
by~\cite{plai-resources-iclp14-short} is 
its use of a 
\emph{sized types} abstract domain, which allows the inference of
non-trivial cost bounds when these depend on the sizes of parts of
input terms 
at any position and depth.
The resulting abstract interpretation-based implementation builds
the cost equations described in
Sect.~\ref{sec:generalization}. 
Separate equations are built for each procedure \emph{version} thanks
to the built-in multivariance in PLAI.
Other optimizations include 
not building equations for unreachable
program parts.

\begin{table}
\footnotesize
\caption{Experimental results (static profiling of accumulated cost).}
\setlength\tabcolsep{1.5pt}
\begin{tabular}{l|l|c|c|c|l|l|c} \hline\hline
\benchmarks
&
\accumcost
&
\compwprev
&
\statvsdyn
&
\atime
&
\stdcost
& 
\numcalls
&
\bigo
\\ \hline\hline

 $\textit{sublist}^*$& $\betatwo+3$ & \multirow{2}{*}{\extendedfeature} 
 & 5\% 
 & 
 4.7 
 &$ \betaone\betatwo+3\betatwo+2 $& 2
 & $\mathcal{O}(\betatwo)$
\\
 $\textit{append}$ & $\betaone\betatwo+2\betatwo-1$ &  
 & $40\%$
 & (NA) & $2\betazero-1$ & $\betaone\betatwo+2\betatwo-1$ & $\mathcal{O}(\betaone\betatwo)$
\\
\hline

 $\textit{is\_prime}^*$ & $1$ &  \multirow{3}{*}{\sameaccuracy} 
 & 0\%
 &  & $ (\variableV - 1)!+\variableV+3  $ & 1 & $\mathcal{O}(1)$
\\
 $\textit{fact}$ & $ \variableV $ &  
 & 0\%
 & 
 1.6
 & $ \variableV  $ &$ \variableV $& $\mathcal{O}(\variableV)$
\\
 $\textit{mult}$ & $ (\variableV - 1)!+2 $ &  
 & 0\%
 & 
 (-24\%)
 & $ \variableV+1  $ & $ (\variableV - 1)!+2 $& $\mathcal{O}(\variableV!)$
 
\\
\hline 

 $\textit{queens}^*$& $ \deltazero+2 $ & \multirow{4}{*}{\extendedfeature} 
 & $7\%$ 
 &  &
 $\mathcal{O}(\deltazero^{\deltazero})^\dagger$ 
 & 1 & $\mathcal{O}(\deltazero)$
\\
 $\textit{consistent}$ & $ \frac{((\deltazero-1) \deltazero-1) \deltazero^{\deltazero+1}+\deltazero}{{(\deltazero-1)}^2}
  $ &
  & $10^4\%$  
 & 
 4.7  
 & $ 2\betazero + 1  $ & $ \frac{((\deltazero-1) \deltazero-1) \deltazero^{\deltazero+1}+\deltazero}{(\deltazero-1)^2}$ & $\mathcal{O}(\deltazero^{\deltazero})$

\\
 $\textit{choose}$ & $ \frac{(2 \deltazero-1) (\deltazero^\deltazero-1)}{(\deltazero-1)}
 $ &  
 & $10^4\%$
 & (NA) & $ 2\deltazero - 1 $ & $\frac{(2 \deltazero-1) (\deltazero^\deltazero-1)}{\deltazero-1}$ 
 & $\mathcal{O}(\deltazero^{\deltazero})$

\\
 $\textit{noattack}$ & $ \frac{(\deltazero-2) \deltazero^{(\deltazero+2)}+\deltazero^2}{(\deltazero-1)^2} $ &  
 & $10^4\%$
 & 
 & $ 1  $ & $\frac{(\deltazero-2) \deltazero^{\deltazero+2}+\deltazero^2}{(\deltazero-1)^2}$ 
 & $\mathcal{O}(\deltazero^{\deltazero})$ 
\\
\hline

 $\textit{search}^*$& $1$ & \multirow{2}{*}{\extendedfeature} 
 & 0\% 
 & 
 1.4 
 &$ 2\betazero+2$& 1 & $\mathcal{O}(1)$
\\
 $\textit{member}$ & $2\betazero+1$ &  
 &0.1\%
 & (NA) & $2\betazero+1$ & $2\betazero + 1$
 & $\mathcal{O}(\betazero)$
\\
\hline

 $\textit{appAll2}^*$ & $b_1$ & \multirow{3}{*}{\sameaccuracy} 
 &$0\%$
 &
 5.3
 & 
 $\mathcal{O}(b_1 b_2 b_3)^\dagger$
 &$1$
 & $\mathcal{O}(b_1)$
\\
 $\textit{appAll}$ & $b_1 b_2$ &  
 &0\% 
 & 
 (-16\%)
 &$b_1 b_2$& $b_1$ & $\mathcal{O}(b_1b_2)$
\\
 $\textit{append}$ & $2 b_1 b_2 b_3$ &  
 & 0\% 
 & &$\betazero$&$b_1 b_2 + b_1$ & $\mathcal{O}(b_1b_2b_3)$
\\
\hline

 $\textit{hanoi}^*$ & $2^\variableV-1$ & \multirow{2}{*}{\sameaccuracy} 
 &$0\%$
 &
 1.6
 &\textbf{$2^{\variableV+1}-2$}& $1$  
 & $\mathcal{O}(2^\variableV)$
\\
 $\textit{move}$ & $2^\variableV-1$ &  
 &$0\%$
 & 
 (-19\%)
  & 1& $2^\variableV-1$ & $\mathcal{O}(2^\variableV)$
\\
\hline

 $\textit{coupled}^*$ & $1$ & \multirow{3}{*}{\sameaccuracy} 
 &$0\%$
 & 
 2.4
 &$\variableV + 2$& 1
 & $\mathcal{O}(1)$
\\
 $p$ & $\frac{\variableV}{2}+\frac{(-1)^\variableV}{4}+\frac{3}{4}$ &  
 &$1.2\%$
 & (-14\%) &$\variableV+1$& $\frac{\variableV}{2}-\frac{(-1)^\variableV}{4}+\frac{1}{4}$ & $\mathcal{O}(\variableV)$
\\
 $q$ & $\frac{\variableV}{2}-\frac{(-1)^\variableV}{4}+\frac{1}{4}$ &  
 &$0\%$
 &  &$\variableV+1$& $\frac{\variableV}{2}+\frac{(-1)^\variableV}{4}-\frac{1}{4}$ &  $\mathcal{O}(\variableV)$
\\
\hline

 $\textit{isort}^*$ & $\betazero+1$ & \multirow{2}{*}{\sameaccuracy} 
 &$0\%$
 & 
 3
 &$\betazero^2+\betazero+1$ & $\betazero+1$
 & $\mathcal{O}(\betazero)$
\\
 $\textit{insert}$ & $\betazero^2$ &  
 &$71\%$
 & 
 (-19\%) 
 &$2\betazero+1$& $\betazero^2$ & $\mathcal{O}(\betazero^2)$
\\
\hline

 $\textit{minsort}^*$ & $\betazero+1$ & \multirow{2}{*}{\sameaccuracy} 
 &$0\%$
 & 
 3.5 
 &$\frac{(\betazero+1)^2}{2}+\frac{\betazero+1}{2}$ & $1$
 & $\mathcal{O}(\betazero)$
\\
 $\textit{findmin}$ & $\frac{(\betazero+1)^2}{2}+\frac{\betazero-1}{2}$ &  
 &$7\%$
 & 
 (-27\%) 
 &$\betazero$& $\betazero+1$ & $\mathcal{O}(\betazero^2)$
\\
\hline

 $\textit{dyade}^*$ & $\betaone$ & \multirow{2}{*}{\sameaccuracy} 
 &$0\%$
 & 
 3.2 
  &$\betaone (\betatwo+1)$&1
 & $\mathcal{O}(\betaone)$
\\
 $\textit{mult}$ & $\betaone \betatwo$ &  
 &$0\%$ 
 &  
 (-20\%)
 & $\betazero$& $\betaone$ & $\mathcal{O}(\betaone\betatwo)$
\\
\hline

 $\textit{variance}^*$ & $1$ & \multirow{3}{*}{\sameaccuracy} 
 &$0\%$
 & 
 3.6 
 &$2\betazero^2$&$1$
 & $\mathcal{O}(1)$
\\
 $\textit{sq\_diff}$ & $\betazero-1$ &  
 &$0\%$ 
 & 
 (-39\%)
  &$2\betatwo\betaone-2\betatwo$&$\betazero-1$ & $\mathcal{O}(\betazero)$
\\
 $\textit{mean}$ & $2\betazero^{2}-\betazero$ &  
 &$0\%$
 & &$2\betazero+1$&$\betazero$ &$\mathcal{O}(\betazero^2)$
\\
\hline

 $\textit{variance2}^*$ & $1$ & \multirow{3}{*}{\sameaccuracy} 
 &$0\%$
 & 
 3.1 
 &$5\betazero +3$& 1 
 & $\mathcal{O}(1)$
\\
 $\textit{sq\_diff}$ & $\betazero$ &  
 &$0\%$
 & 
 (-40\%)
  & $\betazero$  & $\betazero$ & $\mathcal{O}(\betazero)$
\\
 $\textit{mean}$ & $4\betazero + 2$ &  
 &$0\%$
 & &$2\betazero +1$& 2 & $\mathcal{O}(\betazero)$
\\
\hline

 $\textit{listfact}^*$& $b_1$ & \multirow{2}{*}{\sameaccuracy} 
 &$0\%$
 & 
 1.9 
 &$ b_1(b_2+2)$& $1$
 & $\mathcal{O}(b1)$
\\
 $\textit{fact}$ & $b_1b_2+b_1$ &  
 &0\% 
 & 
 (-23\%) 
 &$n$ & $b_1$ & $\mathcal{O}(b_1b_2)$
\\ 
\hline

\end{tabular}
\label{table:ExpResults}
\begin{minipage}{\textwidth}  
\scriptsize 
$\dagger$ For space limitations only the complexity order is shown.

\begin{itemize}
\item $n_1, n_2, \ldots, n_k$ represent the sizes of $k$ input
  arguments. For a single input argument, the subscript is dropped.
\item $b_1, b_2, \ldots, b_k$ represent the sizes of the nested
  structures of an input argument, where $b_1$ represents the size of
  the outer most structure and $b_k$ the inner most. In cases, where
  the cost only depends on the outer most structure, the previous
  representation is used.

\end{itemize}

\end{minipage}
\vspace{-8mm}
\end{table}
Table~\ref{table:ExpResults} shows the results of the comparison
between the proposed approach 
and our previous, program transformation-based approach~\cite{staticprofiling-flops-short}
--\newapproach{} and 
\previousapproach{}
respectively from now on.
Column~\benchmarks{} shows, for each program, the entry predicate
(marked with a \emph{star}, e.g., $sublist^*$) and the predicates
that are declared as cost centers (which always include the entry
predicate).
\accumcost{} shows the parametric
accumulated cost functions inferred for each cost center, which depend
on the input data sizes of the entry predicate.  
For 
conciseness, we only show upper bound
functions, although in the experiments both upper and lower bounds
were inferred. 
The resource
inferred in these tests is the number of resolution steps
(i.e., each clause body is assumed to have unitary cost).
The symbols in Column~\compwprev{} compare \newapproach{} and
\previousapproach{}: 
\extendedfeature{} means that
it is a non-deterministic program that produces multiple solutions
and \newapproach{} is able to obtain non-trivial bounds
while \previousapproach{} fails to obtain a correct bound
(as mentioned before, \previousapproach{} is not applicable for these
programs).\sameaccuracy{}
indicates that \newapproach{} obtains 
the same bounds as \previousapproach{}. Only these two symbols are
required because 
all the results coincide except for the non-deterministic programs.
\textbf{AvD} is the average deviation of
the accumulated costs obtained by evaluating the functions in
Column~\textbf{Acc. Cost}, with respect to the costs measured with a
dynamic profiler~\cite{profiling-padl11-short}. The input data for
dynamic profiling was selected to
exhibit the worst case execution,\footnote{Except for $queens$: the
  \emph{queens} program was simply run for 8 queens. The selection of
  input data that can make a program exhibit worst case execution is
  non-trivial.} in order to compare with upper bound functions.
\atime{} lists the analysis times of \newapproach{} in
\timeunitsfull{} (\ciao/\ciaopp~version 1.15-4048-g6bd1569, 
MacBook Pro,
2.4GHz Intel Core i7 CPU, 8 GB 1333 MHz DDR3 memory, 
MAC OS X Lion 10.7.5)
and, between brackets, how efficient \newapproach{} is with respect to
\previousapproach{} (
$\frac{\newapproach{}-\previousapproach{}}{\previousapproach{}} \times
100)$. \newapproach{} is more efficient than
\previousapproach{} in all
programs, 
with one exception (hanoi).
Times are quite encouraging in any case,
specially considering the currently
inefficient
implementation of the 
Mathematica interface,
one of the solvers used for
the recurrence equations.

\stdcost{} shows the cost functions inferred using the
standard notion of cost (in particular, the cost functions inferred
by~\cite{plai-resources-iclp14-short}) for comparison with the
accumulated cost functions 
(\accumcost). The latter clearly signal hot spots that are not visible
from the standard cost functions. Note also that in all cases the sum
of the functions for all the cost centers is the standard cost of the
entry predicate.  Due to space limitations we do not include analysis
times for obtaining the standard costs in Column~\stdcost{}, but while
the analysis times of \newapproach{} are higher, as 
expected, it is only by $20\%$ on average.
\numcalls{} shows 
the number of times each predicate is called, as a function of input
data sizes of the entry predicate. These functions are inferred using
the standard analysis by defining explicitly a \numcalls{}
\emph{resource} for each cost center predicate.  A large complexity
order in the number of calls to a predicate (in relation to that of a
single call) suggests that it could be profitable to optimize the
program to reduce the number of calls to this predicate,
to 
effectively reduce its impact on the overall cost of the program.
More interestingly,
since both resources \accumcost{} and \numcalls{} of a predicate
\emph{\predp} are expressed as functions of input data sizes of the
entry predicate, their quotient (\accumcost / \numcalls{}) is
meaningful and will give an approximation of the cost of a single call
to $\predp$ as a function of the input data sizes \emph{of the entry
  predicate}.  Note that the standard analysis (Column~\stdcost) also
provides an upper-bound approximation of this cost but as a function
of the input data sizes of
$\predq$.
Finally, Column~\bigo{} shows the \emph{actual} asymptotic resource
usage of the accumulated cost in different cost centers.

\secbeg
\section{Conclusions}
\secend
\label{sec:conclusions}

We
presented a novel, general, and flexible framework for setting up cost
equations/relations which can be instantiated for performing a wide
range of resource usage analyses, including both 
\emph{accumulated cost} 
and 
standard 
cost.
We have also reported on an implementation of this general framework
within the \ciaopp system, and its instantiation for accumulated cost,
and provided some experimental results.
The results show that
the resulting accumulated cost analysis, in addition to
providing results for non-deterministic 
programs, is
also more efficient than our previous approach based on program
transformation, and has a good number of additional advantages.
We argue that our approach is quite general, and it can be easily
applied to other paradigms, including imperative programs, functional
programs, CHR, etc., using the strategy based on compilation to Horn
Clauses, as in our previous work with Java or XC.

\begin{small}
\secbeg

\end{small}

\ifcsname appendices\endcsname
  \appendix
  \clearpage 
  \vspace*{-20mm}
  \centerline{{\LARGE \textbf{Appendices}\footnote{ In the version of
        this paper published in TPLP these appendices constitute the
        supplementary, on-line material associated with the paper. }}}

  \ \\ [-3mm]
  
\secbeg
\section{\\Additional Examples}
\label{sec:additional-examples}
\secend

\begin{example}
\label{examp:sublist}
Consider
the following program to determine whether a list is a sublist of
another.  A sublist can be specified in terms of prefixes and
suffixes: a suffix of a prefix, or a prefix of a suffix. The following
program uses the latter to implement the $\predsublist$ predicate.

\prettylstformat
\begin{lstlisting}
sublist(L, L).
sublist(Sub, List) :- suffix(List, Suf), prefix(Suf, Sub).

suffix(List, Suffix):- append(_, Suffix, List).

prefix(List, Prefix):- append(Prefix, _, List).

append([], L, L).
append([X|Xs], L, [X|Zs]):- append(Xs, L, Zs).
\end{lstlisting}

\noindent

Assume we are going to perform the analysis, to infer upper bounds on
both the standard and accumulated costs, in terms of resolution steps,
for the calling pattern \texttt{sublist(list, list)}, i.e., for the
case where $\predsublist$ is called with both of its arguments bound
to lists.  Given such calling pattern, \ciaopp{} infers the unique
calling patterns \texttt{suffix(list, var)} and \texttt{prefix(var,
  list)}, for $\predsuffix$ and $\predprefix$, where \texttt{var}
represents an unbound variable. However, two different calling
patterns are inferred for $\predappend$:
\texttt{append(var,var,list)}, when it is called from $\predsuffix$,
and \texttt{append(list,var,list)}, when it is called from
$\predprefix$.

Assume that size relations have been inferred for the different
arguments in a clause, and that the size metric used is the \emph{list
  length} of an argument, since all arguments are lists.  The size of
the output (second) argument of $\predsuffix$ is inferred as a
function on 
its input (first) argument, and is represented by
$Sz_{\predsuffix}^{2}(n)$. Such inference sets up the following size
relations:
\begin{equation*}
    \begin{array}{rll}
      Sz_{\predsuffix}^{2}(n) = & Sz_{\predappend}^{2}(n) &  \\ \\ [-3mm]
      
      Sz_{\predappend}^{2}(n) = & 1 &  \text{if } n = 1 \\ 
      Sz_{\predappend}^{2}(n) = & n & \text{if } n > 1 \\ 
      Sz_{\predappend}^{2}(n) = & Sz_{\predappend}^{2}(n-1) & \text{if } n > 1 \\
    \end{array}
\end{equation*}

\noindent
and finds the closed form $Sz_{\predsuffix}^{2}(n)=n$.

In order to infer the standard cost of this program, 
the analysis sets up the following \costrelations{}
for $\predsublist$, $\predsuffix$, $\predprefix$ and
$\predappend$:
\begin{equation*}
    \begin{array}{rll}
      \stdub{\predsublist}{n,m} = & \stdub{\predsuffix}{n} + Sol_{\predsuffix}(n) *  \stdub{\predprefix}{Sz_{\predsuffix}^{2}(n),m} +2 &  \\ 
      \stdub{\predsuffix}{n} = & \stdub{\predappend}{n} + 1 &  \\ 
      \stdub{\predprefix}{n,m} = & \stdub{\predappend}{n,m} + 1  & \\
\end{array}
\end{equation*}

\noindent
Note that the size of the input to the call to $\predprefix$ is given
by the size of the output of $\predsuffix$, represented by
$Sz_{\predsuffix}^{2}(n)$.

The \costrelations{} for the two variants of $\predappend$ are:
\begin{equation*}
    \begin{array}{rll}      
      \stdub{\predappend}{n} = & 1 &  \text{ if } n = 1 \\ 
      \stdub{\predappend}{n} = & \stdub{\predappend}{n-1} + 2  & \text{ if } n > 1 \\ \\
      
      \stdub{\predappend}{n,m} = & 1 &  \text{ if } m = 1, n = 1 \\ 
      \stdub{\predappend}{n,m} = & 1 &  \text{ if } m = 1  \\ 
      \stdub{\predappend}{n,m} = & \stdub{\predappend}{n-1,m-1} + 1  & \text{ if } m > 1 \\ 
      
    \end{array}
\end{equation*}

\noindent 
and the their closed forms are $\stdub{\predappend}{n}=2 \ n-1$ and $\stdub{\predappend}{n,m}=m$
respectively.

Note that in this program $\predsuffix$ produces multiple
solutions. For each solution of $\predsuffix$ the $\predprefix$
predicate is executed on backtracking.

The \costrelations{} for the inference of the number of solutions are:
\begin{equation*}
    \begin{array}{rll}
      Sol_{\predsuffix}(n) = & Sol_{\predappend}(n) &   \\ \\ [-3mm]
      Sol_{\predappend}(n) = & 1 &  \text{if } n = 1\\ 
      Sol_{\predappend}(n) = & Sol_{\predappend}(n-1) + 1 & \text{if } n > 1  \\
    \end{array}
\end{equation*}

\noindent
and the closed form is $Sol_{\predsuffix}(n)=n$.

After composing all the closed forms, the analysis obtains the
following function representing an upper bound on the resource usage
of all the predicates:
\begin{equation*}
    \begin{array}{rll}
      \stdub{\predsuffix}{n} = & 2 \ n &  \\  
      \stdub{\predprefix}{n} = & m + 1  & \\ 
      \stdub{\predappend}{n} = & 2 \ n-1 &   \\ 
      \stdub{\predappend}{n,m} = & m &   \\ 
      \stdub{\predsublist}{n} = & n \ m + 3 \ n + 2 &   \\
    \end{array}
\end{equation*}

Assume now that we declare $\predsublist$ and $\predappend$ as cost
centers to infer the accumulated costs in them. The \costrelations{}
set up for $\predsublist$ are:
\begin{equation*}
    \begin{array}{rll}
      \accub{\predsublist}{\predsublist}{n,m} = & \accube{\predsuffix}{\predsublist}{1}{n}+ Sol_{\predsuffix}(n)* \accube{\predprefix}{\predsublist}{1}{Sz_{\predsuffix}^{2}(n),m} +2 &   \\ \\ [-3mm]

      \accub{\predsublist}{\predappend}{n,m} = & \accube{\predsuffix}{\predappend}{0}{n} + Sol_{\predsuffix}(n)* \accube{\predprefix}{\predappend}{0}{Sz_{\predsuffix}^{2}(n),m} & \\ 
    \end{array}
\end{equation*}

\noindent Replacing the functions for sizes ($Sz$) and solutions ($Sol$) from the previous step we get:
\begin{equation*}
    \begin{array}{rll}
      \accub{\predsublist}{\predsublist}{n,m} = & \accube{\predsuffix}{\predsublist}{1}{n} + n * \accube{\predprefix}{\predsublist}{1}{n,m} + 2 &  \\ [1mm]
      \accub{\predsublist}{\predappend}{n,m} = & \accube{\predsuffix}{\predappend}{0}{n} + n * \accube{\predprefix}{\predappend}{0}{n,m}&  \\ 
    \end{array}
\end{equation*}

\noindent Furthermore, the intermediate cost relations are set up as follows:
\begin{equation*}
    \begin{array}{rll}
      \accube{\predsuffix}{\predsublist}{1}{n} = & \accub{\predappend}{\predsublist}{n} + 1 & \\ 
                                     = & 1 & \text{(Lemma  3)} \\ 
                                     
      \accube{\predprefix}{\predsublist}{1}{n,m} = & \accub{\predappend}{\predsublist}{n,m} + 1 & \\ 
                                     = & 1 & \text{(Lemma  3)} \\ 

      \accube{\predsuffix}{\predappend}{0}{n} = & \accub{\predappend}{\predappend}{n}&   \\ [1mm]

      \accube{\predprefix}{\predappend}{0}{n,m} = & \accub{\predappend}{\predappend}{n,m} &   \\ [1mm]
    \end{array}
\end{equation*}

\begin{equation*}
    \begin{array}{rll}
      \accub{\predappend}{\predappend}{n} = & \accub{\predappend}{\predappend}{n-1} + 2 & \text{if } n > 1   \\ [1mm]
      \accub{\predappend}{\predappend}{n} = & 1 & \text{if } n = 1   \\ [1mm]
    \end{array}
\end{equation*}
\begin{equation*}
    \begin{array}{rll}    
      \accub{\predappend}{\predappend}{n,m} = & 1 & \text{if } m = 1, n = 1   \\ [1mm]
      \accub{\predappend}{\predappend}{n,m} = & 1 & \text{if } m = 1   \\ [1mm]
     \accub{\predappend}{\predappend}{n,m} = & \accub{\predappend}{\predappend}{n-1} + 1 & \text{if } m > 1   \\ [1mm]
      
    \end{array}
\end{equation*}

\noindent
After composing all the intermediate cost relations, the analysis obtains
the following functions representing upper bounds on the accumulated 
resource usage of $\predsublist$ and $\predappend$:
\begin{equation*}
    \begin{array}{rll}
      \accub{\predsublist}{\predsublist}{n} = & n+3 &  \\ \\ [-3mm]
      
      \accub{\predsublist}{\predappend}{n} = & nm+2n-1 &  \\ 
    \end{array}
\end{equation*}
\end{example}

\begin{example}
\label{examp:accumulated-cost}
Consider the following
program $\Prog$:

\prettylstformat
\begin{lstlisting}
p(X, Y) :- i1, i2, q(X, Z),  s(Z, Y).

q(0, 0).
q(X, Y) :- X1 is X - 1, q(X1, Z), i1, i3, s(Z, W), Y is W + 1. 

s(0, 0):- i1.
s(X, Y) :- i2, i4, X1 is X - 1, s(X1, Z), Y is Z + 1.
\end{lstlisting}

Assume that {\tt im}, for {\tt m} $\in \{1, 2, 3,4\}$, are builtin/library
predicates
and that their standard costs
are given by means of trust assertions: for simplicity we assume that
$\Psi({\tt im}) = \stdubna{{\tt im}} = 1$, for {\tt m} $\in \{1, 2, 3,4\}$, 
and that the (standard) cost of the $is/2$ arithmetic predicate is given as zero.
Assume also that $\headcost(\predp)=0 \text{ for all predicates }
\predp \in \Prog$.

Assume that for all the predicates, the first argument is an input
argument and the second one is output, and that the type of all
arguments is the set of natural numbers. Assume that the following
size relationships, expressing the size of the output argument as a
function of the size of the input argument, have already been inferred
for all of them:

\begin{itemize}

\item 
$Sz_{\predp}^{2}(n) = n$, which means that the size (under the natural
  value metric) of the second argument of predicate $\predp$ is
$n$, the size of the input argument.

\item Similarly,
$Sz_{\predq}^{2}(n) = n$, 
and $Sz_{\preds}^{2}(n) = n$.

\end{itemize}

Assume that the set of cost centers is $\ccenters = \{ \predp, \predq,
\preds\}$, and that we want to estimate (upper bounds on) the cost
accumulated in all the cost centers for the predicates $\predp$,
$\predq$, and $\preds$.
Let $\accub{\predp}{\predq}{\vecx}$ denote an upper bound on the
accumulated cost in cost center $\predq$ corresponding to a call
$\predp(\vecx)$.

Assume we process each strongly-connected component of the call graph
of the program in reverse topological order.  We start by inferring
the costs accumulated in cost center $\preds$.  The accumulated cost
in $\preds$ corresponding to a call to $\preds$ is expressed by the
following \costrelation:
\begin{equation*}
\begin{array}{rcl} 
\accub{\preds}{\preds}{0} & = & \accubena{{\tt i1}}{\preds}{1} = \stdubna{{\tt i1}} = 1
\\ \accub{\preds}{\preds}{n} & = & \accubena{{\tt i2}}{\preds}{1} +
\accubena{{\tt i4}}{\preds}{1} + \accub{\preds}{\preds}{n-1} \\
\end{array}
\end{equation*}

\noindent
which can be written as:
\begin{equation*}
\begin{array}{rcl} 
\accub{\preds}{\preds}{0} & = & 1 \\ 
\accub{\preds}{\preds}{n} & = &  2 + \accub{\preds}{\preds}{n-1} \\
\end{array}
\end{equation*}

\noindent
and whose closed-form solution is:
\begin{equation*}
\begin{array}{rcl} 
\accub{\preds}{\preds}{n} & = & 2 \ n + 1. \\
\end{array}
\end{equation*}

Now, we analyze predicate $\predq$. To this end, the accumulated cost
in $\preds$ for a call to $\predq$ is expressed by:
\begin{equation*}
\begin{array}{rcl} 
\accub{\predq}{\preds}{0} & = & 0 \\ 
\accub{\predq}{\preds}{n} & = & \accub{\predq}{\preds}{n-1} + \accubena{{\tt i1}}{\preds}{0} + \accubena{{\tt i3}}{\preds}{0} + \accub{\preds}{\preds}{n-1} \\
\end{array}
\end{equation*}

\noindent
Since there is a trust assertion providing the cost of {\tt i1},
$\Psi({\tt i1}) = 1$ (as already said), according to
Expression~\ref{eq:trustcost},
we have that $\accubena{{\tt i1}}{\preds}{0} = \headcostbool({\tt
  i1},\preds, 0) \times \Psi({\tt i1}) = 0 \times 1 = 0$. Note that
$\headcostbool({\tt i1},\preds, 0) = 0$ because {\tt i1} $\neq$ \preds \ 
and the environment (third argument of $\headcostbool$) is $0$ (since
{\tt i1} is called in the scope of cost center \predq, not in the
scope of the cost center where the analysis is accumulating costs in
this equation, i.e., \preds).
Thus, the cost of {\tt i1} is not taken into account in this equation.
The same consideration applies to {\tt i3}.

Since the cost function for $\accub{\preds}{\preds}{n}$ has already been computed,
replacing values we have
\begin{equation*}
\begin{array}{rcl} 
\accub{\predq}{\preds}{0} & = & 0 \\ 
\accub{\predq}{\preds}{n} & = & \accub{\predq}{\preds}{n-1} + 2 \ (n-1) + 1 \\
\end{array}
\end{equation*}

\noindent
and
\begin{equation*}
\begin{array}{rcl} 
\accub{\predq}{\preds}{0} & = & 0 \\ 
\accub{\predq}{\preds}{n} & = & \accub{\predq}{\preds}{n-1} + 2 \ n - 1 \\
\end{array}
\end{equation*}

\noindent
The solution to the \costrelation{} above is:
\begin{equation*}
\begin{array}{rcl} 
\accub{\predq}{\preds}{n} & = & n^2.\\
\end{array}
\end{equation*}

\noindent
We now analyze predicate $\predp$, so that the accumulated cost in $\preds$
for a call to $\predp$ is expressed by:
\begin{equation*}
\begin{array}{rcl} 
\accub{\predp}{\preds}{n} & = & \accubena{{\tt i1}}{\preds}{0} +
\accubena{{\tt i2}}{\preds}{0} + \accub{\predq}{\preds}{n} +
\accub{\preds}{\preds}{n} \\
\end{array}
\end{equation*}

\noindent
For the same considerations as before, the costs of {\tt i1} and {\tt
  i2} in the body of the clause defining $\predp$ are not taken into
account (i.e., $\accubena{{\tt i1}}{\preds}{0} = \accubena{{\tt
    i2}}{\preds}{0} = 0$).
Replacing values we have that:
\begin{equation*}
\begin{array}{rcl} 
\accub{\predp}{\preds}{n} & = & n^2 + 2n + 1. \\
\end{array}
\end{equation*}

The inference of the accumulated cost in $\preds$ for predicates
$\predp$, $\predq$, and $\preds$ has finished, and we start now the
inference of the accumulated costs in $\predq$.  By
Lemma~\ref{lem:zerocost}, $\accub{\preds}{\predq}{n} = 0$, i.e., we do
not need to analyze predicate $\preds$, since it does not call
$\predq$.  However, now the costs of {\tt i1} and {\tt i3} in the
body of the second clause of $\predq$ do have to be taken into
account.  To this end, the recurrence equations expressing the
accumulated cost in $\predq$ for a call to $\predq$ are:
\begin{equation*}
\begin{array}{rcl} 
\accub{\predq}{\predq}{0} & = & 0 \\ 
\accub{\predq}{\predq}{n} & = & \accub{\predq}{\predq}{n-1} + \accubena{\predq}{{\tt i1}}{1} +
\accubena{\predq}{{\tt i3}}{1}.
\end{array}
\end{equation*}

\noindent
\begin{equation*}
\begin{array}{rcl} 
\accub{\predq}{\predq}{0} & = & 0 \\ 
\accub{\predq}{\predq}{n} & = & \accub{\predq}{\predq}{n-1} + 2. 
\end{array}
\end{equation*}

\noindent
The solution to the recurrence above is:
\begin{equation*}
\begin{array}{rcl} 
\accub{\predq}{\predq}{n} & = & 2 \ n. \\
\end{array}
\end{equation*}

\noindent
Now, the accumulated cost in $\predq$ for a call to $\predp$ is
expressed as:
\begin{equation*}
\begin{array}{rcl} 
\accub{\predp}{\predq}{n} & = & \accub{\predq}{\predq}{n} + \accub{\preds}{\predq}{n} \\
\end{array}
\end{equation*}

\noindent
Replacing values we have that:
\begin{equation*}
\begin{array}{rcl} 
\accub{\predp}{\predq}{n} & = & 2 \ n \\
\end{array}
\end{equation*}

Let us compute now the accumulated cost in $\predp$.  Since it is not
called from $\predq$ nor $\preds$, we have that
$\accub{\predq}{\predp}{n} = \accub{\preds}{\predp}{n} = 0$. The
accumulated cost in $\predp$ for a call to $\predp$ is just the cost
of ${\tt i1}$ and ${\tt i2}$:
\begin{equation*}
\begin{array}{rcl} 
\accub{\predp}{\predp}{n} & = & \accubena{{\tt i1}}{\predp}{1} + \accubena{{\tt i2}}{\predp}{1} = 
\stdubna{{\tt i1}} + \stdubna{{\tt i2}}. 
\end{array}
\end{equation*}

\noindent
Thus:
\begin{equation*}
\begin{array}{rcl} 
\accub{\predp}{\predp}{n} & = & 2. 
\end{array}
\end{equation*}

Note that the \emph{standard cost} of $\predp$ ($\stdub{\predp}{n}$) can be
expressed in terms of the accumulated costs in each of the cost centers:
\begin{equation*}
\begin{array}{rcl} 
\stdub{\predp}{n} & = & \accub{\predp}{\predp}{n} +
\accub{\predp}{\predq}{n} + \accub{\predp}{\preds}{n}.
\end{array}
\end{equation*}
\end{example}

\begin{example}
\label{examp:coupled}

Consider the following program where the predicates $p$ and $q$ are mutually
recursive.

\prettylstformat
\begin{lstlisting}
coupled(X, Y):-	f(X, Y).

p(0,[]).
p(N,[a|R]) :- N1 is N-1, q(N1,R).

q(0,[]).
q(N,[a|R]) :-N1 is N-1, p(N1,R).
\end{lstlisting}

\noindent
Assuming that we want to infer the standard cost of this program in
terms of resolution steps, the analysis sets up the following
\costrelations{} for $coupled$, $p$ and $q$, we
have the following \costrelations:
\begin{equation*}
    \begin{array}{rll}
      \stdub{p}{n} = & 1 &  \text{ if } n = 0 \\
      \stdub{p}{n} = & 1 + \stdub{q}{n-1} & \text{ if } n > 0 \\
    \end{array}
\end{equation*}
\begin{equation*}
    \begin{array}{rll}
      \stdub{q}{n} = & 1 &  \text{ if } n = 0 \\
      \stdub{q}{n} = & 1 + \stdub{p}{n-1} & \text{ if } n > 1 \\
    \end{array}
\end{equation*}
\begin{equation*}
    \begin{array}{rll}
      \stdub{coupled}{n} & = & 1 + \stdub{p}{n}   \\
    \end{array}
\end{equation*}

\noindent 
After composing the closed forms, the analysis obtains 
the following function representing an upper bound on the resource usage of
$coupled$, $p$ and $q$:
\begin{equation*}
    \begin{array}{rll}
      \stdub{coupled}{n} = & n+2 &   \\
    \end{array}
\end{equation*}
\begin{equation*}
    \begin{array}{rll}
      \stdub{p}{n} = & n+1 &   \\
    \end{array}
\end{equation*}
\begin{equation*}
    \begin{array}{rll}
      \stdub{q}{n} = & n+1 &   \\
    \end{array}
\end{equation*}
\noindent

Notice that in this program the cost relations for $p$ and $q$ are
mutually recursive (i.e., they are defined in terms of each other), 
and for this reason the cost functions representing 
the upper bound on the resolution steps in the two are same ($n+1$). Hence, 
the cost of each mutually-recursive predicate subsumes the cost of the other.
However, this cost is in fact distributed between the $p$ and $q$ predicates. 
In order to identify the cost that each of these predicates contributes to
this $n+1$ expression and to the overall cost of $coupled$ ($n+2$), 
we perform the accumulated cost analysis, declaring all the 
predicates as cost centers. 
The instantiation of the equational framework described in
Sect.~\ref{sec:accumulated-cost-inst} obtains the following
accumulated costs for $coupled$, $p$, and $q$:
\begin{equation*}
    \begin{array}{rll}
      \accub{coupled}{coupled}{n} = & 1 &   \\ \\ [-2mm]
      \accub{coupled}{p}{n} = & \frac{n}{2}+\frac{(-1)^n}{4}+\frac{3}{4} &   \\ \\ [-2mm]
      \accub{coupled}{q}{n} = & \frac{n}{2}-\frac{(-1)^n}{4}+\frac{1}{4} &   \\
    \end{array}
\end{equation*}

\noindent
It is now clear how much cost each of $coupled$, $p$, and $q$ contributes
to the standard cost of the whole program ($n+2$). Note that the standard cost
of the mutually recursive predicates $p$ and $q$, which is $n+1$, is now halved among
the two as accumulated costs of $p$ and $q$.

In this example we have shown a hypothetical scenario highlighting that the
accumulated cost information is more useful for mutually recursive parts
of a program in order to identify how much each of the mutually recursive predicates
contributes to the overall cost. This was not possible using only the
standard cost information. 
\end{example}

\begin{example}
\label{examp:evenodd}
Consider the following program to determine the parity of a number where predicates $even$ and $odd$ are mutually recursive.

\prettylstformat
\begin{lstlisting}
even(0).
even(N):- N > 0, N1 is N - 1, odd(N1).

odd(1).
odd(N):- N > 1, N1 is N - 1, even(N1).
\end{lstlisting}

\noindent Similar to the Example~\ref{examp:coupled}, this program contains mutually recursive 
predicates $even$ and $odd$. Since both are defined in terms of each other, the standard analysis 
obtains a same cost function for them representing an upper bound on the resource usage.
\begin{equation*}
    \begin{array}{rll}
      \stdub{even}{n} = & n+1 &   \\
    \end{array}
\end{equation*}
\begin{equation*}
    \begin{array}{rll}
      \stdub{odd}{n} = & n+1 &   \\
    \end{array}
\end{equation*}

In order to identify the cost that each of these predicates contributes to
the overall cost of the program $n+1$, we perform the accumulated cost analysis, 
declaring both $even$ and $odd$ as as cost centers. 
The instantiation of the equational framework described in
Sect.~\ref{sec:accumulated-cost-inst} obtains the following
accumulated costs for $even$ and $odd$:
\begin{equation*}
    \begin{array}{rll}
      \accub{even}{even}{n} = & \frac{n}{2}+\frac{(-1)^n}{4}+\frac{3}{4} &   \\ \\
      \accub{even}{odd}{n} = & \frac{n}{2}-\frac{(-1)^n}{4}+\frac{1}{4} &   \\
    \end{array}
\end{equation*}
\noindent
\end{example}

\vfill
\section{\\Additional Comments on the Relation of the Standard and Accumulated Cost}
\label{sec:std-vs-accum}

Assume that predicate $\predp$ is a cost center.  As already said, in
this case the standard cost of a single call $\predp(\vecx)$
is the
sum of its accumulated costs in all the cost centers in the program.
This is formalized by Theorem 1 in~\cite{staticprofiling-flops-short},
which holds under the assumption that $\predp$ is a cost
center. Intuitively, predicate $\predc$ is ``reachable'' from
predicate $\predp$ if $\predc = \predp$ or $\predc$ can be invoked
(either directly or indirectly) by $\predp$. If $\predp$ is a cost
center, Theorem 1 also holds if we restrict to the set of cost centers
that are reachable from $\predp$, or to the set of cost centers that
are descendants (in the call stack) of $\predp$.  The reason is that
if $\predp$ is a cost center, and another cost center $\predc$
(different from $\predp$) is not reachable from $\predp$, then no part
of the cost of a call to $\predp$ is attributed to $\predc$. This is
stated in Lemma~\ref{lem:zerocost}.

Assume that $\predp$ is the main (entry) predicate in a program, and
that we are interested in knowing how its total (standard) cost is
\emph{distributed} over the (user-defined) cost centers. In this case,
$\predp$ should be declared as a cost center. This is because if
$\predp$ is a cost center, the residual cost (as defined in
Section~\ref{sec:accumulated-cost-inst}) of the call to the main
predicate will be assigned to $\predp$. Otherwise the residual cost
will be left unassigned to any cost center.
 
If $\predp$ is not a cost center, the standard cost of a single call
$\predp(\vecx)$ is the sum of its accumulated costs in all the cost
centers that are descendants (in the call stack) of $\predp$, plus the
residual cost of that call.


\fi

\label{lastpage}
\end{document}